\documentclass[aps,prl,twocolumn,10pt,superscriptaddress,longbibliography]{revtex4-1}
\usepackage{amsmath}
\usepackage{amsfonts}
\usepackage{amssymb}
\usepackage{mathtools}
\usepackage[colorlinks=true,citecolor=blue,linkcolor=blue,urlcolor=blue]{hyperref}
\usepackage{bm}
\usepackage{times}
\usepackage{cases}
\usepackage{wasysym}
\usepackage[version=4]{mhchem}
\usepackage{graphicx}
\usepackage[caption=false]{subfig}

\begin{document}
\title{Double exchange, itinerant ferromagnetism and topological Hall effect in moir\'{e} heterobilayer}

\author{Haichen Jia}
\thanks{These authors contributed equally to this work.}
\affiliation{Department of Physics and HKU-UCAS Joint Institute for Theoretical and Computational Physics at Hong Kong, The University of Hong Kong, Pokfulam Road, Hong Kong, China}

\author{Bowen Ma}
\thanks{These authors contributed equally to this work.}
\affiliation{Department of Physics and HKU-UCAS Joint Institute for Theoretical and Computational Physics at Hong Kong, The University of Hong Kong, Pokfulam Road, Hong Kong, China}
\affiliation{The University of Hong Kong Shenzhen Institute of Research and Innovation, Shenzhen 518057, China}

\author{Rui Leonard Luo}
\affiliation{Department of Physics and HKU-UCAS Joint Institute for Theoretical and Computational Physics at Hong Kong, The University of Hong Kong, Pokfulam Road, Hong Kong, China}

\author{Gang Chen}
\email{gangchen@hku.hk}
\affiliation{Department of Physics and HKU-UCAS Joint Institute for Theoretical and Computational Physics at Hong Kong, The University of Hong Kong, Pokfulam Road, Hong Kong, China}
\affiliation{The University of Hong Kong Shenzhen Institute of Research and Innovation, Shenzhen 518057, China}
    
\date{\today}
    
\begin{abstract}
Motivated by the recent experiments and the wide tunability on the MoTe$_2$/WSe$_2$ moir\'{e} heterobilayer, 
we consider a physical model to explore the underlying physics for the interplay between the 
itinerant carriers and the local magnetic moments. In the regime where
the MoTe$_2$ is tuned to a triangular lattice Mott insulator and the WSe$_2$ layer 
is doped with the itinerant holes, we invoke the itinerant ferromagnetism from the double exchange 
mechanism for the itinerant holes on the WSe$_2$ layer and the local moments on the MoTe$_2$ layer. 
Together with the antiferromagnetic exchange on the MoTe$_2$ layer, the itinerant 
ferromagnetism generates the scalar spin chirality distribution in the system.  
We further point out the presence of spin-assisted hopping in addition to the Kondo coupling 
between the local spin and the itinerant holes, and demonstrate the topological Hall effect
for the itinerant electrons in the presence of the non-collinear spin configurations.
This work may improve our understanding of the correlated moir\'{e}  
systems and inspire further experimental efforts. 
\end{abstract}

\maketitle

%\emph{Introduction.}---

% general introduction
The discovery of the topological and correlation phenomena in the twisted bilayer graphenes
has created a new trend of exploring the emergent physics from the moir\'{e} 
lattice and bands on these heterostructure-like platforms~\cite{Andrei_2020}. 
In fact, 
the fabrication techniques including pulse laser deposition (PLD) and
the molecular beam epitaxy (MBE) has been used for a long time to create 
the heterostructures with layered materials of different physical properties
and different degrees of freedom, 
and modern techniques have been able to fabricate these interfacial heterostructures
with the atomic layer precision~\cite{PhysRevLett.103.256103,science327,nakagawa2006why}. 
The unprecedented controllability and manipulability 
allows these interfaces and heterostructures to access the physical regime and 
the environment that is absent in the bulk materials 
and and thus host the novel and interesting phenomena~\cite{chakhalian2012whither,hwang2012emergent,thiel2006tunable,ohtomo2004high}. 
Besides the epoch-breaking
integer and fractional
quantum Hall effects in the semiconductor heterostructures~\cite{von1980new,PhysRevLett.48.1559},
the modern representative emergent phenomena at the heterostructures 
include the FeSe/SrTiO$_3$ interfacial superconductivity~\cite{wang2012interface},
the Cr-doped Bi$_2$Se$_3$ quantum anomalous Hall effect~\cite{chang2013experimental}, 
interfacial ferromagnetism and polar catastrophe at the LaAlO$_3$/SrTiO$_3$ heterostructure~\cite{ohtomo2004high,li2011coexistence}
and so on.
These results have enriched the scope of topological and correlation physics,
and many have inspired the ongoing theoretical development and progress.

In recent years, the transition metal dichalcogenides have attracted significant attention due to
the intrinsic spin-orbit coupling, topological bands, valleytronics, valley-selective optics,
Ising superconductivity, and the correlation effects~\cite{manzeli20172d,efimkin2018correlation,cao2018topological}. 
The moir\'{e} engineering of the transition metal dichalcogenide
heterostructures further boosted the field into a new direction~\cite{wang2018emerging,novoselov20162d,geim2013van}. 
%Following theoretical results, the experimental discovery of the fractional quantum Hall states in the twisted MoTe$_2$ bilayers at zero field has sparked the recent interest. 
The twisted MoTe$_2$ homobilayers and the MoTe$_2$/WSe$_2$ heterobilayers have inspired 
the more recent interest~\cite{seyler2019signatures}. 
In addition to the band structure topology and the fractional quantum Hall states 
in the twisted MoTe$_2$ bilayers~\cite{PhysRevLett.122.086402,anderson2023programming,Cai_2023,zeng2023integer,tao2022valleycoherent}, 
the moir\'{e} physics of the MoTe$_2$/WSe$_2$ heterobilayer 
allows more possibilities~\cite{Zhao_2023,kang2022switchable,Xu_2022}. 
 The layer-dependent electron correlation
and the displacement-field-controlled layer electron/hole fillings of
 the MoTe$_2$/WSe$_2$ heterobilayer 
 create the bilayer quantum system with both the itinerant electrons
 and the local moments and allow the study of their self and mutual interactions~\cite{kang2022switchable,Zhao_2023,Xu_2022}.
 In the case where the correlated MoTe$_2$ moir\'{e} lattice is filled, one may simulate the triangular lattice Hubbard model and reveal the physics of Mott transition and Mott magnetism~\cite{Li_2021}. In the case where the itinerant carriers are introduced on the WSe$_2$ layer, the heavy fermion related physics was identified. 
 On top of these important developments, we explore other 
 emergent physics that may realize on this interesting heterostructure platform.

% specific introduction 

\begin{figure}[b]
\includegraphics[width=8cm]{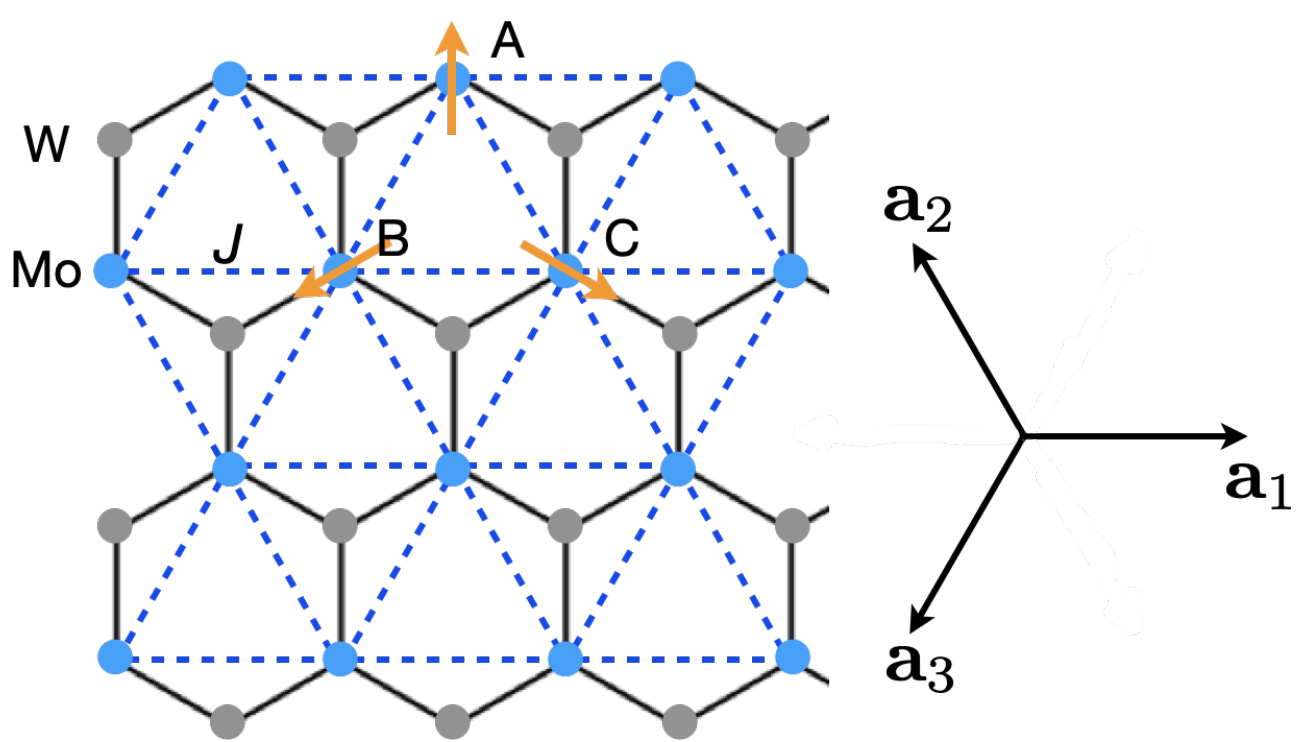}
\caption{The moir\'{e} Kondo bilayer forms a honeycomb lattice. 
The moir\'{e} sites of the MoTe$_2$ (WSe$_2$) layer are labeled by
the blue (gray) color. The dashed line refers to the antiferromagnetic 
exchange coupling between the local moments on the MoTe$_2$ sites. 
Here ${\boldsymbol a}_1$, ${\boldsymbol a}_2$, and ${\boldsymbol a}_3$ are the 
three basis vectors of the triangular bravais lattice. }
\label{fig1}
\end{figure}

%\emph{Double exchange and itinerant ferromagnetism.}---
In the MoTe$_2$/WSe$_2$ moir\'{e} heterobilayer, it was inferred from the experimental results and the DFT
calculations that, the MoTe$_2$ layer is more correlated than the WSe$_2$ layer~\cite{Zhang_2021}. 
This is probably because the wavefunctions of the $5d$ electrons from the W atom are more extended
than the $4d$ electrons from the Mo atom. 
The holes in the MoTe$_2$ layer can be tuned to become local moments, while the holes in the WSe$_2$ layer
can remain itinerant. The heavy fermion type of physics with the Kondo hybridization between the itinerant holes
from the WSe$_2$ layer and the local moments from the MoTe$_2$ layer was observed. To explore the
candidate physics, we consider the following Kondo-lattice-like model,
\begin{eqnarray}
H &=&  J   \sum_{\langle\langle ij \rangle\rangle_{\text{Mo}}}{\boldsymbol S}_i \cdot {\boldsymbol S}_j 
- t\sum_{\langle\langle ij \rangle\rangle_{\text{W}}}  ( c^{\dagger}_{i\sigma} c^{}_{j\sigma} + h.c.)
\nonumber \\
&& + J_K \sum_{\langle ij \rangle} {\boldsymbol S}_i \cdot  (c^{\dagger}_{j\alpha} 
 \frac{{\boldsymbol{\sigma}}_{\alpha\beta}}{2} c^{}_{j\beta} )
+ \cdots ,
\label{eq1}
\end{eqnarray}
where the $J$-term is the exchange interaction between the local moments on the Mo layer, 
the $t$-term is the hole tunneling on the W layer, the $J_K$-term is the Kondo coupling 
between the itinerant hole on the WSe$_2$ layer and the local moment on the MoTe$_2$ layer,
and the ``$\cdots$'' refers to other interactions such as the spin-assisted hopping of the holes 
that will be discussed later. The subindex Mo (W) refers to the MoTe$_2$ (WSe$_2$) layer. 
With the ``$1+x$'' doping, the MoTe$_2$ moir\'{e} sites are occupied by the local magnetic moments,
and the remaining $x$ introduces the $x$ hole doping on the WSe$_2$ moir\'{e} sites. 
As shown in Fig.~\ref{fig1},
 the moir\'{e} lattice of the heterobilayer is a honeycomb lattice. The moir\'{e} sites of 
the MoTe$_2$ layer form one triangular sublattice, and the moir\'{e} sites of the 
WSe$_2$ layer form the other triangular sublattice. The operator
${\boldsymbol S}_i$ refers to the spin-1/2 local moment on the MoTe$_2$ moir\'{e} sites,
and $c^{\dagger}_{i\alpha}$ ($c^{}_{i\alpha}$) creates (annihilates) a hole 
with spin $\alpha$ at the moir\'{e} lattice site $i$ on the WSe$_2$
layer.  

Although the model in Eq.~\eqref{eq1} is written with the $t$ and $J$ terms,
the physics should differ from the $t$-$J$ model for the doped Hubbard model where the
correlation is uniform throughout the system. For the heterobilayer here, at the current stage, it seems that,
the strong Hubbard interaction only occurs on the MoTe$_2$ moir\'{e} sites, but not 
on the WSe$_2$ moir\'{e} sites. Unlike the $t$-$J$ model where the hole and the spin exchange
each other during the hopping, the hole-hopping only happens on the WSe$_2$ layer. 
Therefore, if one attempts to make the connection with the existing physical contexts, 
the more appropriate one is probably the double exchange model for the doped
 manganite perovskites with mixed valences of the Mn
ions~\cite{PhysRev.82.403,1960Gennes,PhysRevLett.118.027203}. Despite the connection, there still exist some key differences. 
In the doped manganites, the itinerant carriers from the upper $e_g$ orbitals 
are on the same lattice sites as the local moments
from the lower t$_{2g}$ orbitals, and they interact with a ferromagnetic Hund's coupling instead 
of an antiferromagnetic Kondo coupling. 
Moreover, the parent model for Eq.~\eqref{eq1} is actually the Anderson model~\cite{hewson_1993},
while the double exchange model is the bare model for the coupling between 
two species of electrons from the $e_g$ and $t_{2g}$ manifolds.

For the double exchange in doped manganites, the local moment from the $t_{2g}$ orbitals tends
to polarize the spin of the itinerant $e_g$ electrons. In the strong Hund's coupling limit, the spinor 
wavefunction of the $e_g$ electron is locked to the local moment of each lattice site. 
When the $e_g$ electron tunnels on the lattice, 
the overlap of the spinor wavefunctions from neighbor sites determines the effective hopping,
and a ferromagnetic spin configuration is preferred 
in order to gain the kinetic energy. When the antiferromagnetic superexchange between the 
$t_{2g}$ moments is considered, a compromised spin configuration with the spin tilting
between the antiferromagnetic order and the collinear ferromagnet is obtained. 
For the model in Eq.~\eqref{eq1}, such a ferromagnetic onsite Hund's coupling
is replaced by the antiferromagnetic Kondo coupling from the neighboring sites. 
Following the spirit of double exchange, the three neighboring spins together 
polarize the spin of the itinerant hole site at the WSe$_2$ layer. 
As the itinerant hole only hops between the moir\'{e} sites on the WSe$_2$ layer,
to gain kinetic energy, 
 a ferromagnetic component in the spin configuration would be preferred. 
 Since the Kondo coupling is not as strong as the Hund's coupling for the manganites,
the spinor-charge separation in Zener's treatment is not directly applicable to the current context
to understand the magnetic structure~\cite{PhysRev.82.403}. 
Instead, we directly consider the total energy of Eq.~\eqref{eq1} for the system 
in the presence of a bit more generic spin configuration.

\begin{figure}[t]
\includegraphics[width=7.5cm]{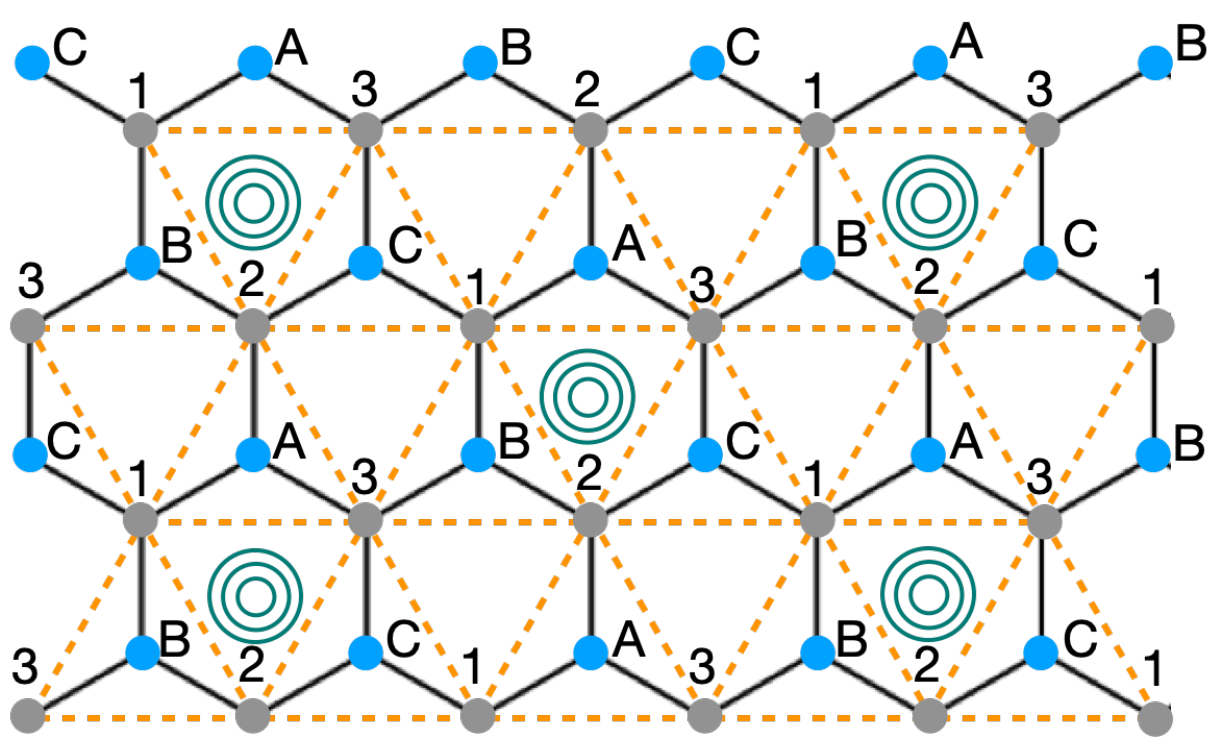}
\caption{The sublattices in the presence of the spin-assisted hopping. ``1,2,3'' refer to the 
sublattices for the itinerant holes. 
The enlarged unit cell is labeled by the hexagon plaquette with the marked circle inside.}
\label{fig2}
\end{figure}

With only the exchange coupling on the MoTe$_2$ layer, the ground state of the local moments
is simply the 120-degree magnetic order with three sublattices (see Fig.~\ref{fig1}). 
As it is expected to have ferromagnetic components, 
we then parameterize the spin configuration on the three magnetic sublattices as 
\begin{eqnarray}
\langle {\boldsymbol S}_{\text{A}} \rangle &=& S (0, \sin \theta, \cos \theta),  \\
 \langle {\boldsymbol S}_{\text{B}} \rangle &=& S \big( {- \sqrt{3} } ( \sin \theta)/2, -( \sin \theta )/2, \cos \theta \big),\\ 
  \langle {\boldsymbol S}_{\text{C}} \rangle &=& S \big(\sqrt{3} (\sin \theta) /2, -(\sin \theta) /2, \cos \theta \big),
\end{eqnarray}
where ${S=1/2}$, ${\theta = 0}$ for the fully polarized state, and ${\theta = \pi/2}$
for the coplanar 120-degree state. One can then single out the itinerant part of 
the model in Eq.~\eqref{eq1} in the presence of the underlying magnetic order, 
and their contribution is given as
\begin{eqnarray}
&&- t\sum_{\langle\langle ij \rangle\rangle_{\text{W}}}  ( c^{\dagger}_{i\sigma} c^{}_{j\sigma} + h.c.) 
+ \frac{3}{2}S J_K \cos \theta \sum_{i \in \text{W}}  c^{\dagger}_{j\alpha}  {{{\sigma}}^z_{\alpha\beta}}  c^{}_{j\beta} 
\nonumber \\
&&= \sum_{\boldsymbol k} 
\epsilon_+  ( {\boldsymbol k} )   c^{\dagger}_{{\boldsymbol k}\uparrow} c^{}_{{\boldsymbol k}\uparrow} 
+
\epsilon_-  ( {\boldsymbol k} )    c^{\dagger}_{{\boldsymbol k}\downarrow} c^{}_{{\boldsymbol k}\downarrow} ,\label{eq5}
\end{eqnarray}
with
\begin{equation}
\epsilon_{\pm} ( {\boldsymbol k} ) =  {-2t \sum_{\mu=1,2,3}  \cos ( {\boldsymbol k} \cdot {\boldsymbol a}_{\mu}) \pm  \frac{3}{2}S J_K \cos \theta }.
\end{equation}
Thus, there are two bands that are separated by the ferromagnetic components. 
In the dilute doping limit, if the hole carriers simply occupy the band bottom 
of the lower band. Including the exchange energy, the total energy of the system is 
\begin{eqnarray}
&& E_{tot} \simeq ( -6t -\frac{3}{2}S J_K \cos \theta ) \frac{x N}{2}  
\nonumber \\
&& \quad\quad\quad\quad\quad\quad +\frac{3NJS^2}{4}(3\cos^2\theta-1),
\end{eqnarray}
where $N$ is the total number of moir\'{e} sites. 
The minimal energy is established for a tilted spin configuration with 
a finite ferromagnetic component, 
\begin{equation}
\cos \theta = \frac{J_K}{6JS} x .\label{theta}
\end{equation}

A model with the basic ingredients in Eq.~\eqref{eq1} is sufficient to generate a 
weak ferromagnetism via the mechanism of the double exchange. With this spin configuration,
the system supports a finite scalar spin chirality within the magnetic unit cell with
\begin{eqnarray}
\chi=\langle ( {\boldsymbol S}_{\text A} \times {\boldsymbol S}_{\text B} ) \cdot {\boldsymbol S}_{\text C} \rangle \neq 0  .
\label{eq9}
\end{eqnarray}
Usually, when a finite scalar spin chirality is present in a system with both local moments and itinerant 
carriers, there should exist a topological Hall effect from the coupling between these two different 
degrees of freedom. The scalar spin chirality generates a real-space Berry flux for the itinerant carriers,
and generates a topological Hall effect by twisting the motion of the carriers. 
This is absent in the present form of Eq.~\eqref{eq1}
as the simple Kondo coupling in Eq.~\eqref{eq1} does not capture this physics.  
The itinerant hole experiences the magnetization of the three sublattices as an average, and 
cannot experience the scalar spin chirality. 
To obtain the topological Hall effect, one has to return to the parent model, i.e., 
the Anderson model~\cite{hewson_1993}. 

The original Anderson model involves the hybridization between the itinerant electron
and the local electron orbitals as well the onsite Hubbard-$U$ interaction that 
creates the local magnetic moment. When one goes from the Anderson model to the
Kondo lattice model via the second-order perturbation, in addition to the Kondo coupling
between the itinerant carriers and the local moments, there exists a spin-assisted hopping 
term for the itinerant carriers. From the minimalist's point of view, we consider the following
shortest range spin-assisted hopping, 
\begin{equation}
\tilde{t} \big[  \sum_{\langle\langle ij \rangle\rangle}
\sum_{k \in (ij)}
 ( c^{\dagger}_{i\alpha} {\boldsymbol{\sigma}}^{}_{\alpha\beta} c^{}_{j\beta} ) \cdot {\boldsymbol S}_{ k} 
+( c^{\dagger}_{j\alpha} {\boldsymbol{\sigma}}^{}_{\alpha\beta} c^{}_{i\beta} ) \cdot {\boldsymbol S}_{ k}  
  \big],
  \label{eq10}
\end{equation}
where $k$ is the local moment site in between the sites $ij$ for the itinerant holes.  
Ideally, it would be better to combine Eq.~\eqref{eq10} and the Eq.~\eqref{eq1},
and carry out a complete analysis. Here, we take the phenomenological treatment
and consider Eq.~\eqref{eq10} as a perturbation to Eq.~\eqref{eq1} with a small $\tilde{t}$ (compared to $J_K$). 
In this approximation, the presence of the spin-assisted hopping in Eq.~\eqref{eq10} 
does not significantly modify the underlying magnetic structure. 

% explain the hopping within the unit cell. 

With the three-sublattice magnetic order, the unit cell is tripled (see Fig.~\ref{fig2}),
and the sublattices for the WSe$_2$ moir\'{e} lattice are labeled as ``1,2,3''. 
To see the effect of spin-assisted hopping, we simply illustrate the point
with one unit cell on the lower left corner of Fig.~\ref{fig2} before performing 
the actual calculation. 
Following the
treatment of the double exchange theory~\cite{PhysRev.82.403}, 
we first express the electron operator
as ${c_{i\sigma} = c_i z_{i\sigma}}$ where $c_i$ is spinless fermion at $i$
and $z_{i\sigma}$ is the spinor representing the spin state of the itinerant hole.
This representation is simply adopted for our physical understanding and cannot be used
for our actual calculation. 
 In this representation, 
the spin-assisted hopping of Eq.~\eqref{eq10} becomes
\begin{equation}
\tilde{t} \sum_{\langle\langle ij \rangle\rangle } \sum_{k \in (ij)}
\big[ (z^{\ast}_{i\alpha} {\boldsymbol{\sigma}}_{\alpha\beta} z_{j\beta} ) c_i^\dagger c_j^{} + 
( z^{\ast}_{j\alpha} {\boldsymbol{\sigma}}_{\alpha\beta} z_{i\beta} ) c_j^\dagger c_i^{}  \big] \cdot {\boldsymbol S}_k.
\end{equation}
If ${z_{i\sigma} = z_{j \sigma}}$, one should have 
$z^{\ast}_{i\alpha} {\boldsymbol{\sigma}}_{\alpha\beta} z_{i\beta}$ 
to be aligned (anti-aligned) with ${\boldsymbol S}_k$ for $\tilde{t}<0$ ($\tilde{t}>0$)
 to optimize the energy. 
When the hole hops from 1 to 2, the spin of the hole tends to be aligned
with the local moment from the B sublattice. 
When the hole hops from 2 to 3, the spin of the hole tends to be aligned
with the local moment from the C sublattice. 
When the hole hops from 3 back to 1, the spin of the hole tends to be aligned
with the local moment from the A sublattice.
 Thus, when the hole 
goes around from 1 to 2 to 3 then to back 1, 
the spin wavefunction of the hole is twisted from being aligned with the A sublattice order to 
the B sublattice, then to the C sublattice, and finally back to the 
A sublattice. Since the spin states are different on the A,B,C sublattices, 
such a uniform choice of $z$ is frustrated to optimize the energy. 
 This is an example of itinerant frustration. 
 A nonuniform spinor $z$ is generically expected for the 1,2,3 sublattices to optimize the 
 spin-assisted hopping. With the nonuniform $z$, the original kinetic energy in Eq.~\eqref{eq5} 
 becomes 
 \begin{equation}
 -t \sum_{\langle\langle ij \rangle\rangle_{\text{W}}} ( z^{\ast}_{i\sigma} z^{}_{j\sigma} c^\dagger_i c^{}_j + h.c.),
 \end{equation}
 where $t z^{\ast}_{i\sigma} z^{}_{j\sigma} $ is an effective complex hopping. The complex
 phase of $t z^{\ast}_{i\sigma} z^{}_{j\sigma} $ can then function
 as an effective U(1) gauge field for the charged spinless fermion. In the context of 
 the double exchange for the 
doped manganites or the magnetic skyrmion lattices for itinerant magnets, 
due to the approximate adiabatic condition from the strong {\sl onsite}
Hund's coupling, 
the spinor is pinned by the onsite magnetic order, and 
the effective U(1) gauge flux is directly related to the solid angle spanned 
by the noncollinear spin configurations~\cite{PhysRevLett.83.3737,Kurumaji_2019}. 
Here, no such relation is obtained 
as spin-assisted hopping is a bond coupling term instead of the onsite term,
and we are also not really in the adiabatic limit. 
Although this current process differs from the spinor alignment of the 
itinerant carriers via the
onsite Kondo/Hund coupling with the local magnetic orders 
in the context of itinerant magnets, 
the physical outcome is rather similar. 
Since the spinor $z_i$ configuration is not expected to be uniform due to the 
itinerant frustration, the effective real-space U(1) gauge flux is still expected here.
Due to the weakness of $\tilde{t}$ and the bond coupling with the spin
instead of the onsite coupling, the actual effective flux experienced by
the hole is certainly different from the usual expectation
from the solid angle spanned by the spins in Eq.~\eqref{eq9}.
Nevertheless, 
the topological Hall effect of the hole carriers is still expected as well
as the resulting momentum 
space Berry curvature distribution.

\begin{figure}[t]
	\includegraphics[width=1.0\linewidth]{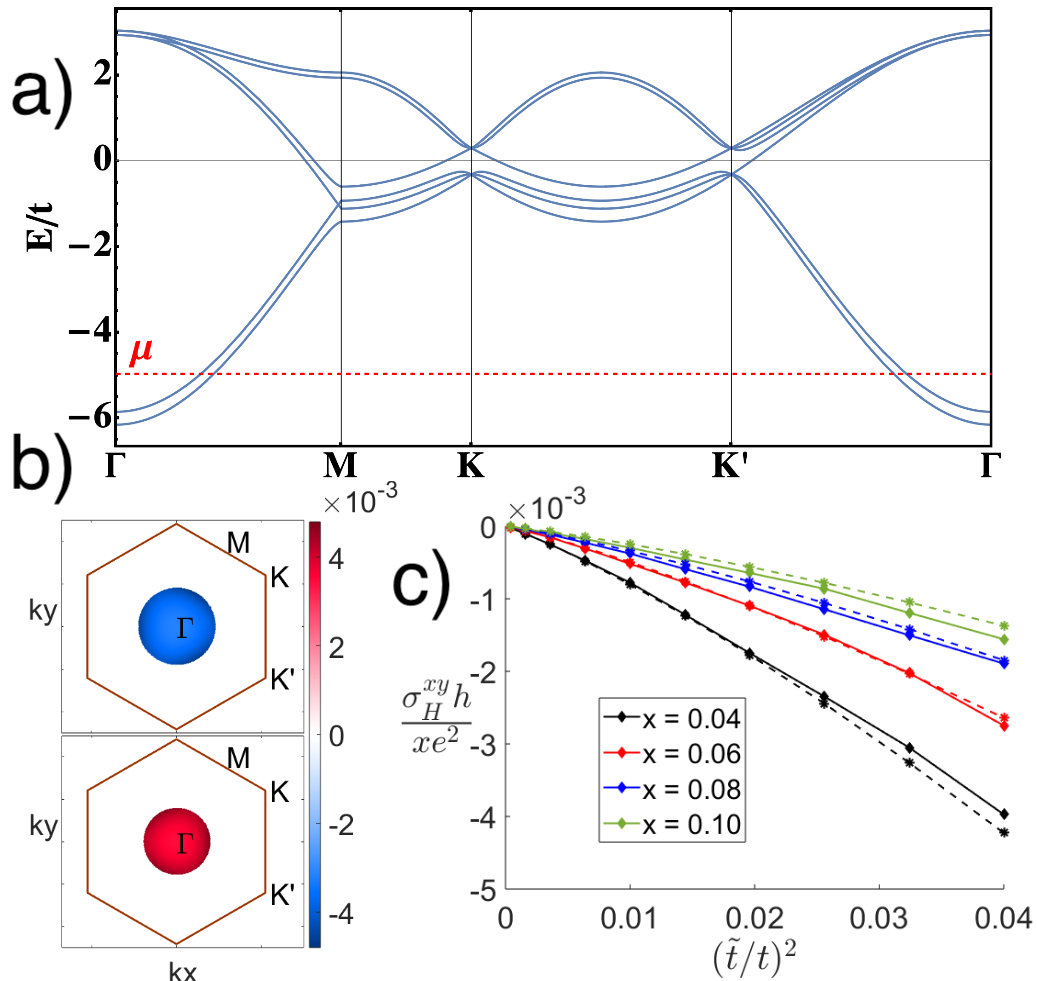}
	\caption{a) The (folded) band dispersion along high symmetry directions. We choose $\tilde{t} = 0.2t$, $J = 0.3t$, $J_K = t$, 
	with doping $x = 0.1$ and $\theta = 83.6^{\circ}$. 
	The Fermi level corresponding to ${x = 0.1}$ is ${\mu = -4.96t}$ (red dotted line).
b) Berry curvature distribution for the states under the Fermi level. Brown hexagon indicates the Brillouin Zone boundary. The upper (lower) part shows the Berry curvature of the lowest (second lowest) band. 
c) The Hall conductance $\sigma_H^{xy}$ for various $\tilde{t}$ and filling $x$, in the unit of ${xe^2}/{h}$. Other parameters are the same as those in (a). Solid lines indicate numerical calculation while dotted lines are computed from analytical estimate Eq.~\eqref{eq15}. }
	\label{fig:fig3}
\end{figure}

In the absence of the spin-assisted hopping, the hole experiences the magnetic 
order via the off-site Kondo coupling, and the effect is equivalent to a uniform Zeeman coupling
because the A,B,C sublattice orders apply to the hole site together and the antiferromagnetic
parts are canceled out. Such a Zeeman coupling does not create nontrivial nor finite Berry curvature distribution
of the itinerant holes. The spin-assisted hopping transfers the effect of the noncollinear spin texture 
to the itinerant holes, such that the hole band acquires the Berry curvature distribution in the 
momentum space. In Fig.~\ref{fig:fig3}, we combine Eq.~\eqref{eq1} and Eq~\eqref{eq10}, and 
explicitly compute dispersion, Berry curvature distribution and Hall conductance for the hole band with the noncollinear spin order as the background.
The Hall conductance is given by the Kubo formula~\cite{kubo1957statistical,PhysRevLett.49.405} as
\begin{align}
    \sigma_H^{xy}&=\frac{e^2}{\hbar}\frac{1}{A}\sum_{n,\boldsymbol{k}}\Theta(\mu-E_{n\boldsymbol{k}})\Omega_{n\boldsymbol{k}},
\end{align}
where $A$ is the area of the $xy$-plane, $\mu$ is the Fermi level determined from doping, 
$\Theta$ is the unit step function as the Fermi-Dirac distribution function at zero-temperature limit, 
$|n\boldsymbol{k}\rangle\ (\langle n\boldsymbol{k}|)$ is the ket (bra)-eigenvector for the $n$-th folded band, 
$E_{n\boldsymbol{k}}$ is its dispersion, and $\Omega_{n\boldsymbol{k}}$ is the $n$-th band Berry curvature. 
After incorporating the three-lattice order, the hole bands are folded into six bands (see Fig.~\ref{fig:fig3}). 
As shown in Fig.~\ref{fig:fig3}, the Fermi level $\mu$ is just above the band bottom in the dilute doping case, 
the finite Berry curvature around $\boldsymbol{\Gamma}$ then indeed leads to a non-zero Hall conductance.

In Eq.~\eqref{theta}, we have concluded that a ferromagnetic angle $\theta$ can emerge from a simplified consideration, 
while the interactions of the real materials can make Eq.~\eqref{theta} more complicated, so in Fig.~\ref{fig:fig3} we instead 
phenomenologically choose a weak ferromagnetic state with $\theta=83.6^{\circ}$, and numerically compute the Hall conductance. 
To explicitly show the effect of the spin-assisted hopping and the spin chirality, in the dilute doping limit, we approximately consider the Berry curvature contribution $\Omega_{1,\boldsymbol{k}=0}\equiv \Omega_1$ and $\Omega_{2,\boldsymbol{k}=0}\equiv \Omega_2$ from the band bottom of the two lowest bands, i.e.,  $E_{1,\boldsymbol{k}=0}\equiv E_1$ and $E_{2,\boldsymbol{k}=0}\equiv E_2$ respectively, so that
\begin{align}
    \sigma_H^{xy}&\approx\frac{e^2}{h}\frac{Nx}{2A}\frac{(\mu-E_1)\Omega_1+(\mu-E_2)\Omega_2}{(\mu-E_1)+(\mu-E_2)}\nonumber\\
    &=\frac{e^2}{h}\left\{\frac{8\pi t(J_K+\tilde{t})}{\left[9t^2-(J_K+\tilde{t})^2S^2\cos^2\theta\right]^2}x\right.\nonumber\\
    &\left.-\frac{2(J_K+4\tilde{t})\left[9t^2+(J_K+\tilde{t})^2S^2\cos^2\theta\right]}{3t\left[9t^2-(J_K+\tilde{t})^2S^2\cos^2\theta\right]^2}\right\}\chi \tilde{t}^2,\label{eq15}
\end{align}
where the details of the calculation can be found in the appendix. 
From this result, it is clear that there is a finite Hall conductance contribution only when the noncollinear 
spin configuration holds a non-zero spin chirality $\chi$ and the Hall conductance is approximately proportional to $\tilde{t}^2$, which is consistent with the numerical results in Fig.~\ref{fig:fig3}.

\emph{Discussion.}---The interplay between the itinerant carriers and 
the local moments in the MoTe$_2$/WSe$_2$ moir\'{e} heterobilayer 
reveals the physics of geometric frustration and itinerant frustration 
that create the noncollinear spin texture and the topological Hall effect
via the double exchange mechanism. The current origin of correlation is
mainly from the MoTe$_2$ layer with the moir\'{e} triangular lattice. 
The MoTe$_2$/WSe$_2$ moir\'{e} heterobilayer has a honeycomb lattice
without the inversion symmetry. 
It will be interesting to increase the correlation on the WSe$_2$ layer as well
such that the system is close to the honeycomb lattice Hubbard
model where the inversion symmetry is absent. Since both the 
interlayer and intralayer coupling/tunneling can be tuned,
the system can simulate many rich physics including 
spiral spin liquid~\cite{PhysRevB.81.214419,Yao_2021}, deconfined quantum criticality~\cite{PhysRevLett.110.127203},
doped Mott insulator~\cite{PhysRevB.88.155112} and so on.

Since it has been demonstrated that, the MoTe$_2$ layer could be tuned
from metal to Mott insulator via the Mott transition by more-or-less varying the correlation. 
Since the antiferromagnetic order appears in the strong Mott regime, 
this immediately raises the fate of the weak Mott regime where the electron correlation is 
in the intermediate range. Two possibilities have been discussed.  
One is the spinon Fermi surface U(1) spin liquid where the spinon sector forms a non-Fermi liquid~\cite{PhysRevLett.95.036403}. 
In this case, the time-reversal symmetry is preserved, and the interesting 
direction is to explore the Kondo effect of this spinon non-Fermi liquid with the itinerant carriers. 
The other one is the chiral spin liquid that was proposed based on the density matrix renormalization
group (DMRG) calculation
and understood from a recent mean-field study~\cite{PhysRevX.10.021042,PhysRevLett.127.087201}. 
In this case, the time reversal is broken and there
is a chiral edge mode that should support quantized thermal Hall conductance~\cite{zhang2023thermal}. 
The coupling of this chiral Abelian topologically ordered state with the itinerant carriers
would still generate the topological Hall effect for the itinerant carriers as the 
scalar spin chirality of the chiral spin liquid can still be experienced by the itinerant carriers. 
Moreover, the charge-neutral chiral edge mode of the chiral spin liquid may be detected electrically by 
placing the $C=1$ chiral edge of integer quantum Hall states on both sides,
and such a resonant-tunneling-like idea for the charge-neutral chiral edge mode
has been proposed for the (chiral) Kitaev spin liquid and $\alpha$-RuCl$_3$~\cite{PhysRevX.10.031014}. 
Here, the MoTe$_2$/WSe$_2$ system may be a bit more experimentally feasible as it can naturally
support the $C=1$ quantum Hall state~\cite{tao2022valleycoherent}.

To summarize, we have explained the itinerant ferromagnetism 
from the perspective of double exchange 
and further predicted the topological Hall effect due to the 
interplay between the itinerant carriers and the local magnetic moments
from the moir\'{e} potentials for the MoTe$_2$/WSe$_2$ heterobilayer. 
In addition to these phenomena due to the magnetic orders, the more exotic 
scenarios were further discussed and visioned.  

\emph{Acknowledgments.}---We acknowledge Kin Fai Mak for the discussion and Wang Yao for the arrangement. 
This work is supported by 
the Ministry of Science and Technology of China with Grants No. 2021YFA1400300, 
the National Science Foundation of China with Grant No. 92065203, 
and by the Research Grants Council of Hong Kong with C7012-21GF. 
\\
%%%%%%%%%% Prefix a "A" to all equations, figures, tables and reset the counter %%%%%%%%%%
\setcounter{equation}{0}
\setcounter{figure}{0}
\setcounter{table}{0}
\setcounter{page}{1}
\makeatletter
\renewcommand{\theequation}{A\arabic{equation}}
\renewcommand{\thefigure}{A\arabic{figure}}
\renewcommand{\citenumfont}[1]{A#1}
%%%%%%%%%% Prefix a "A" to all equations, figures, tables and reset the counter %%%%%%%%%%
\section{Appendix A: Hamiltonian Matrix}
In this section, we consider the effective Hamiltonian for the itinerant holes at W sites with both Kondo coupling $J_K$ and spin-assisted hopping $\tilde{t}$ as 
\begin{align}
    H_\text{W}=&-t\sum_{\langle\langle ij \rangle\rangle_{\text{W}},\sigma}(c_{i\sigma}^\dagger c_{j\sigma}+h.c.)\nonumber\\
    &+J_K\sum_{\langle ij \rangle,\alpha\beta}\boldsymbol{S}_i\cdot\left(c^\dagger_{j\alpha}\frac{\boldsymbol{\sigma}_{\alpha\beta}}{2}c_{j\beta}\right)\nonumber\\
    &+\tilde{t}\left[\sum_{\langle\langle ij \rangle\rangle_\text{W}}\sum_{k\in(ij)}\sum_{\alpha\beta}\left(c^\dagger_{i\alpha}\boldsymbol{\sigma}_{\alpha\beta}c_{j\beta}+h.c.\right)\cdot\boldsymbol{S}_k\right],
\end{align}
where the notations are the same as those in the main text. The W sites form a triangular lattice with the six neighboring bonds as $\pm\boldsymbol{a}_1=\pm(1,0)$, $\pm\boldsymbol{a}_2=\pm(-1/2,\sqrt{3}/2)$ and $\pm\boldsymbol{a}_3=\pm(-1/2,-\sqrt{3}/2)$. Obviously, the non-coplanar ``umbrella'' spin ordering and the presence of $\tilde{t}$ terms triple the unit cell with three sublattices denoted as 1, 2 and 3. The basis vectors of the enlarged lattice are then $\boldsymbol{R}_1=(3/2,-\sqrt{3}/2)$ and $\boldsymbol{R}_2=(3/2,\sqrt{3}/2)$. The corresponding reciprocal vectors are $\boldsymbol{b}_1=(2\pi/3,-2\pi\sqrt{3}/3)$ and $\boldsymbol{b}_2=(2\pi/3,2\pi\sqrt{3}/3)$.

By performing the Fourier transform in the enlarged lattice, we obtain the Hamiltonian in the basis of $\Psi_{\boldsymbol{k}}=(c_{1\boldsymbol{k}\uparrow},c_{1\boldsymbol{k}\downarrow},c_{2\boldsymbol{k}\uparrow},c_{2\boldsymbol{k}\downarrow},c_{3\boldsymbol{k}\uparrow},c_{3\boldsymbol{k}\downarrow})^T$ as
\begin{widetext}
    \begin{align}
    H_\text{W}(\boldsymbol{k})=\Psi^\dagger_{\boldsymbol{k}}
    \begin{bmatrix}
        \frac{3}{2}J_K S \cos\theta\ & 0 &f_{+}(\boldsymbol{k}) & g_{3}(\boldsymbol{k}) & f^*_{+}(\boldsymbol{k}) & g_{2}(-\boldsymbol{k})\\
        0 & -\frac{3}{2}J_K S \cos\theta\ & g^*_{3}(-\boldsymbol{k}) & f_{-}(\boldsymbol{k}) & g^*_{2}(\boldsymbol{k}) & f^*_{-}(\boldsymbol{k})\\
        f^*_{+}(\boldsymbol{k}) & g_{3}(-\boldsymbol{k}) & \frac{3}{2}J_K S \cos\theta\ & 0 & f_{+}(\boldsymbol{k}) & g_{1,\boldsymbol{k}}\\
        g^*_{3}(\boldsymbol{k}) & f^*_{-}(\boldsymbol{k}) & 0 & -\frac{3}{2}J_K S \cos\theta\ & g^*_{1}(-\boldsymbol{k}) & f_{-}(\boldsymbol{k})\\
        f_{+}(\boldsymbol{k}) & g_{2}(\boldsymbol{k}) & f^*_{+}(\boldsymbol{k}) & g_{1}(-\boldsymbol{k}) & \frac{3}{2}J_K S \cos\theta\ & 0\\
        g^*_{2}(-\boldsymbol{k}) & f_{-}(\boldsymbol{k}) & g^*_{1,\boldsymbol{k}} & f^*_{-}(\boldsymbol{k}) & 0 & -\frac{3}{2}J_K S \cos\theta\
    \end{bmatrix}\Psi_{\boldsymbol{k}},\label{HW}
\end{align}
\end{widetext}
where $f_\sigma(\boldsymbol{k})=-(t-\sigma \tilde{t} S\cos\theta)\sum_{\mu=1,2,3} e^{-i\boldsymbol{k}\cdot\boldsymbol{a_\mu}}$ and $g_n(\boldsymbol{k})=-i \tilde{t} S\sin\theta \sum_{\mu=1,2,3} e^{-i\left[\boldsymbol{k}\cdot\boldsymbol{a}_\mu-\frac{2\pi}{3}(\mu+n)\right]}$. By diagonalizing the matrix in Eq.~(\ref{HW}), we obtain the band dispersion as shown in Fig.~3 in the main text.
%%%%%%%%%% Prefix a "B" to all equations, figures, tables and reset the counter %%%%%%%%%%
\setcounter{equation}{0}
\setcounter{figure}{0}
\setcounter{table}{0}
\setcounter{page}{1}
\makeatletter
\renewcommand{\theequation}{B\arabic{equation}}
\renewcommand{\thefigure}{B\arabic{figure}}
\renewcommand{\citenumfont}[1]{B#1}
%%%%%%%%%% Prefix a "S" to all equations, figures, tables and reset the counter %%%%%%%%%%
\section{Appendix B: Hall Conductance Estimate}
In this section, we analytically calculate the Hall conductance in the dilute doping limit. In this case, most related physics can be captured in the energy region between the band bottom and the chemical potential $\mu$. Since $t$ and $J_K>0$, when $|\tilde{t}|\ll J_K$, the band bottom occurs at $\boldsymbol{\Gamma}$ point, i.e. $\boldsymbol{k}=(0,0)$. Fortunately, an exact diagonalization of Eq.~(\ref{HW}) can be achieved at $\boldsymbol{\Gamma}$, and the two lowest energies are
\begin{align}
    E_{1,2}=-6t\mp\frac{3}{2}S\cos\theta(J_K+4\tilde{t}).
\end{align}
When $\tilde{t}=0$, the lowest energy $E_1$ recovers Eq.~(6) in the main text, and small $\tilde{t}$ indeed only affects $\theta$ perturbatively.

Moreover, we can analytically calculate the Berry curvature of these two bands at $\boldsymbol{\Gamma}$ as
\begin{align}
    \Omega_{1,2}=\mp\frac{3 S^2 \tilde{t}^2\sin^2\theta}{2\left[3t\pm (J_K+\tilde{t})S\cos\theta\right]^2}.
\end{align}

Approximately,
\begin{align}
    \sigma_H^{xy}&\approx\frac{e^2}{\hbar A_0}\left[\Omega_{1}\int\frac{d^2\boldsymbol{k}}{(2\pi)^2}f(E_{1\boldsymbol{k}})+\Omega_{2}\int\frac{d^2\boldsymbol{k}}{(2\pi)^2}f(E_{2\boldsymbol{k}})\right]\nonumber\\
    &=\frac{e^2}{2\pi h A_0}\left[\Omega_{1}\int_{E_1}^\mu dE\frac{d^2\boldsymbol{k}}{dE_{1\boldsymbol{k}}}+\Omega_{2}\int_{E_2}^\mu dE\frac{d^2\boldsymbol{k}}{dE_{2\boldsymbol{k}}}\right],
\end{align}
where $A_0=\frac{\sqrt{3}}{2}$ is the area per W sites.

Again, since $|\tilde{t}|\ll J_K$, we estimate $\frac{d^2\boldsymbol{k}}{dE_{1(2),\boldsymbol{k}}}$ using the dispersion relation at $\tilde{t}=0$, i.e. $\epsilon_{\pm} ( {\boldsymbol k} )=-2t\sum_\mu\cos(\boldsymbol{k}\cdot\boldsymbol{a}_\mu)\pm\frac{3}{2}J_KS\cos\theta\approx\frac{3}{2}t|\boldsymbol{k}|^2-6t\pm\frac{3}{2}J_KS\cos\theta$, to get $\frac{d^2\boldsymbol{k}}{dE_{1(2),\boldsymbol{k}}}\approx\frac{d^2\boldsymbol{k}}{d\epsilon_{\pm,\boldsymbol{k}}}\approx\frac{2\pi}{3t}$.

We can also estimate $\mu$ from $x$-doping as
\begin{align}
    x&=\int_{E_1}^\mu \frac{dE}{(2\pi)^2}\frac{d^2\boldsymbol{k}}{dE_{1\boldsymbol{k}}}+\int_{E_2}^\mu \frac{dE}{(2\pi)^2}\frac{d^2\boldsymbol{k}}{dE_{2\boldsymbol{k}}}\nonumber\\
    &\approx\frac{\mu-E_1}{6\pi t}+\frac{\mu-E_2}{6\pi t}=\frac{\mu+6t}{3\pi t},
\end{align}
or alternatively,
\begin{align}
    \mu=3(\pi x-2)t
\end{align}
Then,
\begin{widetext}
    \begin{align}
    \sigma_H^{xy}&\approx\frac{e^2}{h A_0}\left(\Omega_{1}\frac{\mu-E_1}{3t}+\Omega_{2}\frac{\mu-E_2}{3t}\right)\nonumber\\
    &=\frac{e^2}{h}\sqrt{3}S^2\tilde{t}^2\sin^2\theta\left\{\frac{\pi x-\frac{(J_K+4\tilde{t})S\cos\theta}{2t}}{\left[3t-(J_K+\tilde{t}) S\cos\theta\right]^2}-\frac{\pi x+\frac{(J_K+4\tilde{t})S\cos\theta}{2t}}{\left[3t+(J_K+\tilde{t}) S\cos\theta\right]^2}\right\}\nonumber\\
    &=\frac{e^2}{h}
    \frac{12 \pi t^2 (J_K+\tilde{t})x-(J_K+4\tilde{t})\left[9t^2+(J_K+\tilde{t})^2S^2\cos^2\theta\right]}{t\left[9t^2-(J_K+\tilde{t})^2S^2\cos^2\theta\right]^2}\frac{2}{3}\chi \tilde{t}^2
\end{align}
\end{widetext}
which is proportional to spin chirality $\chi$ and $\tilde{t}^2$.
\bibliography{refs}
\end{document}